\documentclass[journal,comsoc,twocolumn]{IEEEtran}

\usepackage[utf8]{inputenc}

\usepackage[T1]{fontenc}
\pdfoutput=1
\usepackage{tikz}
\usetikzlibrary{positioning}
\usepackage{amsmath, amssymb}
\usepackage{graphicx}
\usepackage{subcaption}
\usepackage{autobreak}
\allowdisplaybreaks

\usepackage{float}
\usepackage{enumerate}

\usepackage{bm}

\usepackage{cleveref}

\usepackage{setspace}

\usepackage{color}
\usepackage{multirow}
\usepackage{upgreek}

\usepackage[integrals]{wasysym}
\usepackage{ulem}

\title{
All-fiber photonic lantern multimode optical receiver with coherent adaptive optics beam combining}

\author{Bo Zhang,
Jianfeng Sun,
Chenzhe Lao,
Christopher H Betters, 
Alexander Argyros,
Yu Zhou\\
and
Sergio Leon-Saval, ~\IEEEmembership{Member,~IEEE,}
\thanks{Bo Zhang, Jianfeng Sun, Chenzhe Lao  and  Yu Zhou are with Key Laboratory of Space Laser Communication and Detection Technology, Shanghai Institute of Optics and Fine Mechanics, Chinese Academy of Sciences, 390 Qinghe Rd., Shanghai 201800, China; and Center of Materials Science and Optoelectronics Engineering, University of Chinese Academy of Sciences, Beijing 100049, China
(e-mail: zhangbo@siom.ac.cn, sunjianfengs@163.com, 251728782@qq.com and sunny@mail.siom.ac.cn).}
\thanks{Christopher H Betters and Sergio Leon-Saval are with Sydney Astrophotonic Instrumentation Laboratory, School of Physics, The University of
Sydney, NSW 2006, Australia
(e-mail: christopher.betters@sydney.edu.au, 
sergio.leon-saval@sydney.edu.au).}
\thanks{Alexander Argyros is with Matrix Education, Sydney, NSW 2000, Australia
(e-mail: aarg4123@gmail.com).}}

\begin{document}

\maketitle

\begin{abstract}
A multimode optical receiver for free space optical communications (FSOC) based on a photonic lantern and adaptive optics coherent beam combining (CBC) of the lantern's single-mode outputs is proposed and demonstrated for the first time. The use of optical coherent combining in fiber serves to increase the signal to noise ratio compared to similar receivers based on electrically combined signals, and represents an all-fiber approach to low-order adaptive optics. This optical receiver is demonstrated using a photonic lantern with three outputs, fibre couplers and active phase locking, and further investigated under atmospheric conditions with and without turbulence.   
\end{abstract}


\section{Introduction}

Lately, free space optical communications (FSOC) have been widely studied as a disruptive alternative for large capacity communications in satellite and terrestrial laser links. When using multiple ground stations although, simplified optical systems are desirable to allow the usage of more flexible approaches. In terrestrial and space-to-ground laser communications and receivers, several methods for optical system simplification are indeed possible. For instance, one way of reducing the required power-aperture product on a space platform is to  implement  effective,  but  costly,  single-aperture  ground  terminals  with  large  collection areas, e.g. large telescope apertures. Another approach recently demonstrated for increasing  collection area and efficiency is to combine signals from multiple apertures using digital coherent combination (DCC)\cite{Geisler:16}. Thus, using many small less-expensive apertures combined to create a large effective aperture while maintaining good receiver sensitivity. However, the multi-aperture nature of the process imposes other constraints and requirements on the digitization. by 

\begin{figure}[t]
\centering
\includegraphics[width=0.48\textwidth, draft=false]{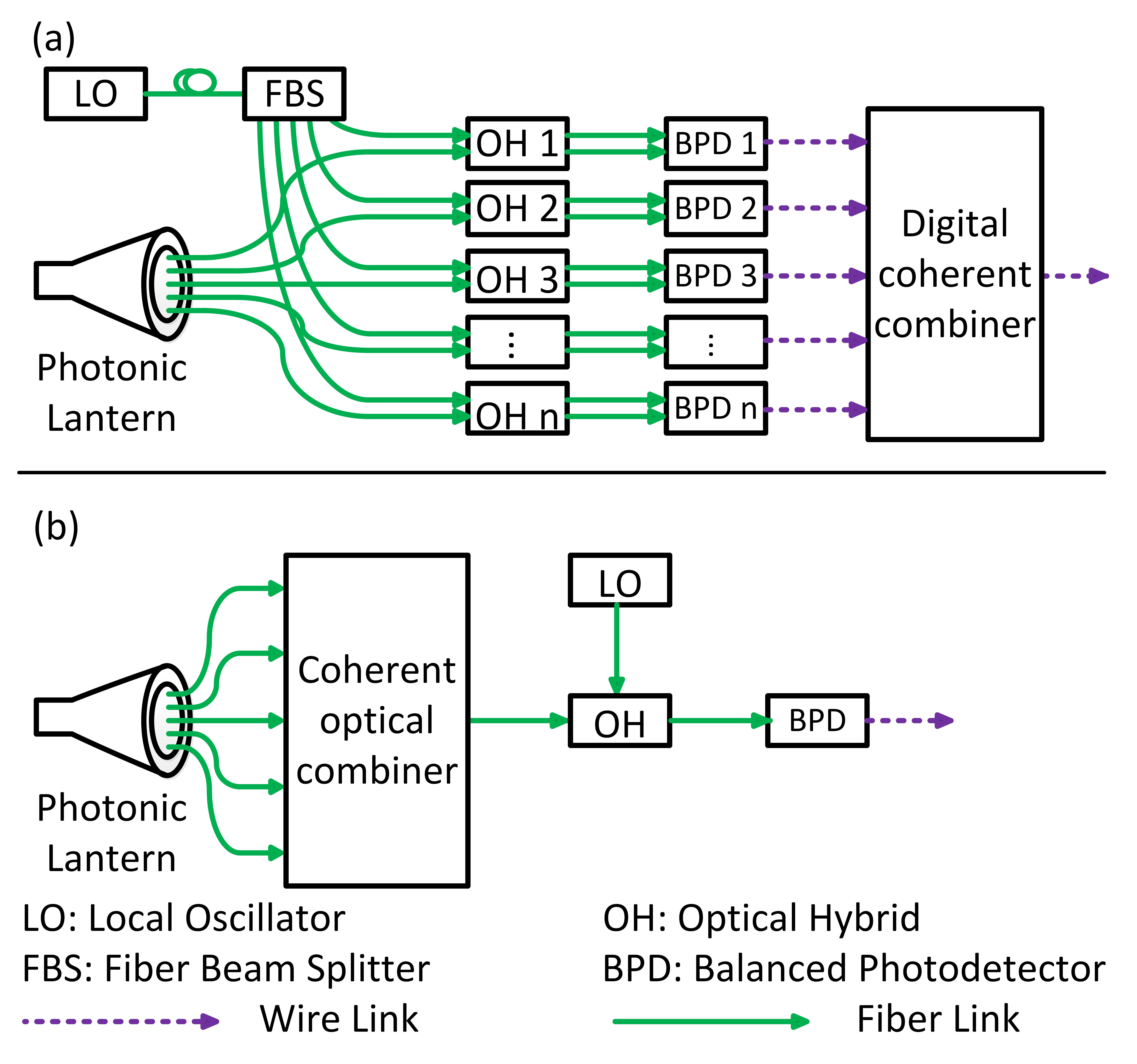}
\caption{Structural diagram of the PL-based optical receiver: (a) using the method of digital coherent combination; (b) using the method of coherent beam combining. }
\label{fig: PL}
\end{figure}

Recently, a multimode single-aperture optical receiver was proposed \cite{BOZHANG2019, Ozdur2015} based on photonic lanterns \cite{Birks2015, Leon2013, Leon2005} and the direct coherent combination of signals in the electrical domain. The proposed scheme is shown in Fig.~\ref{fig: PL}a and involves a photonic lantern converting a multimode signal into $N$ single-mode signals, which are mixed with single-mode local oscillator beams in optical hybrids. The mixed signals are then read by $N$ balanced photodiodes and coherently combined electronically. This scheme takes advantage of the high coupling efficiency of a multimode signal to the multimode end of a photonic lantern \cite{ Ozdur2013} and the higher tolerance in free-space coupling situations to tilt errors \cite{Yarnall2017}. It also takes advantage of the high mixing efficiency achieved in single-mode fibre (SMF) \cite{Jacob1995}, and has been used in laser detection and ranging (LIDAR) \cite{ Ozdur2015} and free-space optical communications \cite{Zheng2018}.

However, the above method may result in lowering the signal-to-noise ratio (SNR). This is because the original multimode signal is divided into $N$ parts, meaning the signal at each photodiode is reduced on average by a factor of $N$. Hence the SNR at each photodiode is reduced, as is the average over all photodiodes. Depending on the circumstances, the reduction of SNR using this DCC approach may negate the advantages of using the photonic lantern in the first place, as compared to directly coupling to one single-mode fibre and photodiode. If the input is not diffraction limited (i.e. multimode, due to atmospheric turbulence), then the lantern will be able to increase the signal through a higher coupling efficiency by capturing high order modes not guided by the SMF. However, this is only an advantage if the increase in signal compensates for the reduction in SNR. If the input is already diffraction limited (i.e. single-mode, in the absence of atmospheric turbulence and tilt and tip errors), the lantern provides no advantage in terms of coupling \cite{zheng2016} but may reduce SNR.  

Here, we propose a photonic lantern-based optical receiver that avoids the reduction in SNR described above by using a single photodiode and an adaptive optics in-fiber coherent beam combining (CBC), i.e. coherently combining the signals optically in optical fiber couplers rather than electronically \cite{YanYang2017multi}. The receiver is shown in Fig. \ref{fig: PL}b, and the key difference is that the $N$ single-mode outputs of the lantern are combined into one single-mode beam using a coherent optical combiner. The beam is then mixed with the local oscillator and detected by a single photodiode. This retains the high coupling efficiency of the lantern and high mixing efficiency and long propagation distance bandwidth of SMFs, and eliminates the decrease in SNR that arises from dividing the signal between multiple photodiodes. 

In this paper we demonstrate the operation of the CBC-based receiver described. We verify the feasibility of optically combining the outputs using adaptive optics based on active phase locking and coherent interferometry in 3-dB fiber couplers. In addition, we use a phase screen to verify the feasibility of the optical receiver under conditions of atmospheric turbulence.

\section{Coherent beam combining}

\subsection{Principle of operation}
\label{section: Principle}

Photonic lanterns with adaptive spatial mode control have been recently demonstrated to selectively excite only the fundamental mode in large mode area amplifiers  \cite{Montoya2016, MontoyaJuan2017}. To demonstrate our proposed photonic lantern-based receiver with the coherent combination of beams in the optical domain we applied an active phase locking architecture on the single-mode fiber pigtails combined with 3-dB fiber couplers. Thus,  relatively simple feedback loops from detectors to adaptive optic phase elements allowed the device to adapt automatically to any specific input beam form \cite{MillerDavidAB2013} while coherently combining all single-mode output power into one single-mode fiber.  A lantern with three single-mode outputs was used, the schematic and microscope image are shown in Fig.~\ref{fig: SETUP}. The laser signal beam passes through Lens 1 and then a phase screen to simulate atmospheric turbulence. Lens 2 couples the signal into the multimode end of the photonic lantern which transitions adiabatically to three single-mode fibres at the output. 

The first and second output fibres are connected to a 3-dB coupler (Coupler-1 in the Fig. \ref{fig: SETUP}) and the phase of the light in the second fibre is controlled by the phase shifter PS-1, which is controlled by the photodiode PD-1 that measures the second output of the coupler. Minimising the reading on PD-1 results in the first and second outputs being combined coherently. This combined output is then combined coherently with the third output from the lantern using the same method with PS-2 and PD-2.

\subsection{Theoretical evaluation of efficiency in the system}
\label{section: Theoretical derivation}

In order to get a better understanding of the optical receiver works in this optical coherent beam combiner architecture, it is important to evaluate the phase, optical losses and power output relationships between the different components. To this purpose we model the efficiency of the all-fiber CBC receiver as follows. The transfer matrix of the $k $th 3-dB fiber coupler is
\begin{equation}
{{M}_{k}}=\sqrt{\frac{\zeta_{k}}{2}} \left[ \begin{matrix}
 1 & j \\
 j & 1 \\
\end{matrix} \right], 
\ \ k=1, 2
\label{eq: M3dB}
\end{equation}

\noindent
where ${\zeta_{k}}$ is the insertion loss of the $k$th coupler.

If we take the single-mode outputs of the $1 \times 3$ lantern to be ${{P}_{S, 1}}$, ${{P}_{S, 2}}$ and ${ {P}_{S, 3}}$, and their phases to be ${{\theta }_{S, 1}}$, ${{\theta }_{S, 2}}$ and ${{\theta }_{S, 3}}$, 
the electric field outputs (${{E}_{O, 1}}$ and ${{E}_{O, 2}}$) of the first 3-dB coupler (Coupler 1)  can be expressed by
\begin{equation}
\left[ \begin{matrix}
 {{E}_{O, 1}} \\
 {{E}_{O, 2}} \\
\end{matrix} \right]={{M}_{1}}
\left[ \begin{matrix}
 \exp \left( -j\frac{\Delta {{\theta }_{12}}}{2} \right) & 0 \\
 0 & \exp \left( j\frac{\Delta {{\theta }_{12}}}{2} \right) \\
\end{matrix} \right]
\left[ \begin{matrix}
 \sqrt{{{P}_{S, 1}}} \\
 \sqrt{ \xi_{1} {{P}_{S, 2}}} \\
\end{matrix} \right]
\label{eq: E12}
\end{equation}

\noindent
where $\Delta {{\theta }_{12}}={{\theta }_{S, 1}}-{{\theta }_{S, 2}}$ is the phase difference between the output of the single-mode fibers 1 and 2, and $\xi_{1} $ is the loss factor of the first phase shifter (PS-1) used to control the phase difference between fiber 1 and 2 before the first 3-dB coupler (Coupler-1).  
The first output of Coupler-1 and the third output of the lantern (fibre 3) pass through the second 3-dB fiber coupler (Coupler-2). Thus, the electric field outputs of the two arms of Coupler-2 (${{E}_{O, 3}}$ and ${{E}_{O, 4}}$) can be given by
\begin{equation}
\left[ \begin{matrix}
 {{E}_{O, 3}} \\
 {{E}_{O, 4}} \\
\end{matrix} \right]
={{M}_{2}}
\left[ \begin{matrix}
 \exp \left( -j\frac{\Delta {{\theta }_{123}}}{2} \right) & 0 \\
 0 & \exp \left( -j\frac{\Delta {{\theta }_{123}}}{2} \right) \\
\end{matrix} \right]
\left[ \begin{matrix}
 \sqrt{{\xi}_{2}} |{{E}_{O, 1}}| \\
 \sqrt{{{P}_{S, 3}}} \\
\end{matrix} \right]
\label{eq: E34}
\end{equation}

\noindent
where $\Delta {{\theta }_{123}}={{\theta }_{O, 1}}-{{\theta }_{S, 3}}$, is the phase difference between the first output of the Coupler-1 (${{P}_{O, 1}}$) and the fibre 3 output of the lantern (${{P}_{S, 3}}$); and where ${{\theta }_{O, 1} }$ is the phase value of ${{P}_{O, 1}}$, and $\xi_{2} $ is the loss factor of the second phase shifter (PS-2). Hence, the intensity of each output can be obtained by the magnitude of the electric field squared:

\begin{equation}
\left\{ \begin{aligned}
 & {{P}_{O, 1}}=\frac{\zeta_{1}}{2}
 \left[
 {P}_{S, 1}+\xi_{1} {P}_{S, 2}-2 \sqrt{ \xi_{1} {P}_{S, 1} {P}_{S, 2}} \sin ({\Delta {{\theta }_{12}}})
 \right ]\\
 & {{P}_{O, 2}}=\frac{\zeta_{1}}{2}
 \left[
 {P}_{S, 1}+\xi_{1} {P}_{S, 2}+2 \sqrt{ \xi_{1} {P}_{S, 1} {P}_{S, 2}} \sin ({\Delta {{\theta }_{12}}})
 \right ]\\
 & {{P}_{O, 3}}=\frac{\zeta_{2}}{2}
 \left[
 {\xi_{2}}{P}_{O,1}+{P}_{S, 3}-2 \sqrt{{\xi_{2}}{P}_{O,1} {P}_{S, 3}} \sin ({\Delta {{\theta }_{123}}})
 \right ]\\
 & {{P}_{O, 4}}=\frac{\zeta_{2}}{2}
 \left[
 {\xi_{2}}{P}_{O,1}+{P}_{S, 3}+2 \sqrt{{\xi_{2}}{P}_{O,1} {P}_{S, 3}} \sin ({\Delta {{\theta }_{123}}})
 \right ].
\end{aligned} \right. 
\label{eq: Icohere1}
\end{equation}

The second outputs of the two couplers (${{P}_{O, 2}}$ and ${{P}_{O, 4}}$) are directed to the photodiodes. When the phase differences between the two inputs of each coupler are $\frac{3}{2} \pi +2 \pi n $, for $n$ an integer, the first outputs of each coupler will be maximum achieving coherent optical combination into one single-mode fiber output. This gives
\begin{equation}
\left\{ 
\begin{aligned}
 & {{P}_{O, 1}}=\frac{\zeta_{1}}{2}{{\left( \sqrt{{P}_{S, 1} }+\sqrt{\xi_{1} {P}_{S, 2}} \right)}^{2}} \\
 & {{P}_{O, 2}}=\frac{\zeta_{1}}{2}{{\left( \sqrt{{P}_{S, 1} }-\sqrt{\xi_{1} {P}_{S, 2}} \right)}^{2}} \\
 & {{P}_{O, 3}}=\frac{\zeta_{2}}{2}
 {{\left[ \sqrt{\frac{\zeta_{1} \xi_{2}{{P}_{S, 1}}}{2}}
 +\sqrt{\frac{\zeta_{1} \xi_{1} \xi_{2} {P}_{S, 2}}{2}}
 +\sqrt{{{P}_{S, 3}}} \right]}^{2}} \\
 & {{P}_{O, 4}}=\frac{\zeta_{2}}{2}
 {{\left[ \sqrt{\frac{\zeta_{1} \xi_{2}{{P}_{S, 1}}}{2}}
 +\sqrt{\frac{\zeta_{1} \xi_{1} \xi_{2} {P}_{S, 2}}{2}}
 -\sqrt{{{P}_{S, 3}}}
 \right]}^{2}}. 
\end{aligned} 
\right.
\label{eq: Icohere2}
\end{equation}

 Incidentally, ${{P}_{O, 2}}$ and ${{P}_{O, 4}}$ are used to control the phase of the first and second phase shifters (PS-1 and PS-2) respectively, and ${{P}_{O, 3}}$ is the final coherently combined single-mode signal output.
From this evaluation it can be stated that the coherent synthesis efficiency ($\eta $) is thus
\begin{equation}
\eta =\frac{{{P}_{O, 3}}}{{{P}_{S, 1}}+{{P}_{S, 2}}+{{P}_{S, 3}}}.
\label{eq: effi}
\end{equation}

\section{Experimental work}
\label{section: Experiments}

\begin{figure*}
\begin{center}
\includegraphics [width=0.9\textwidth, draft=false] {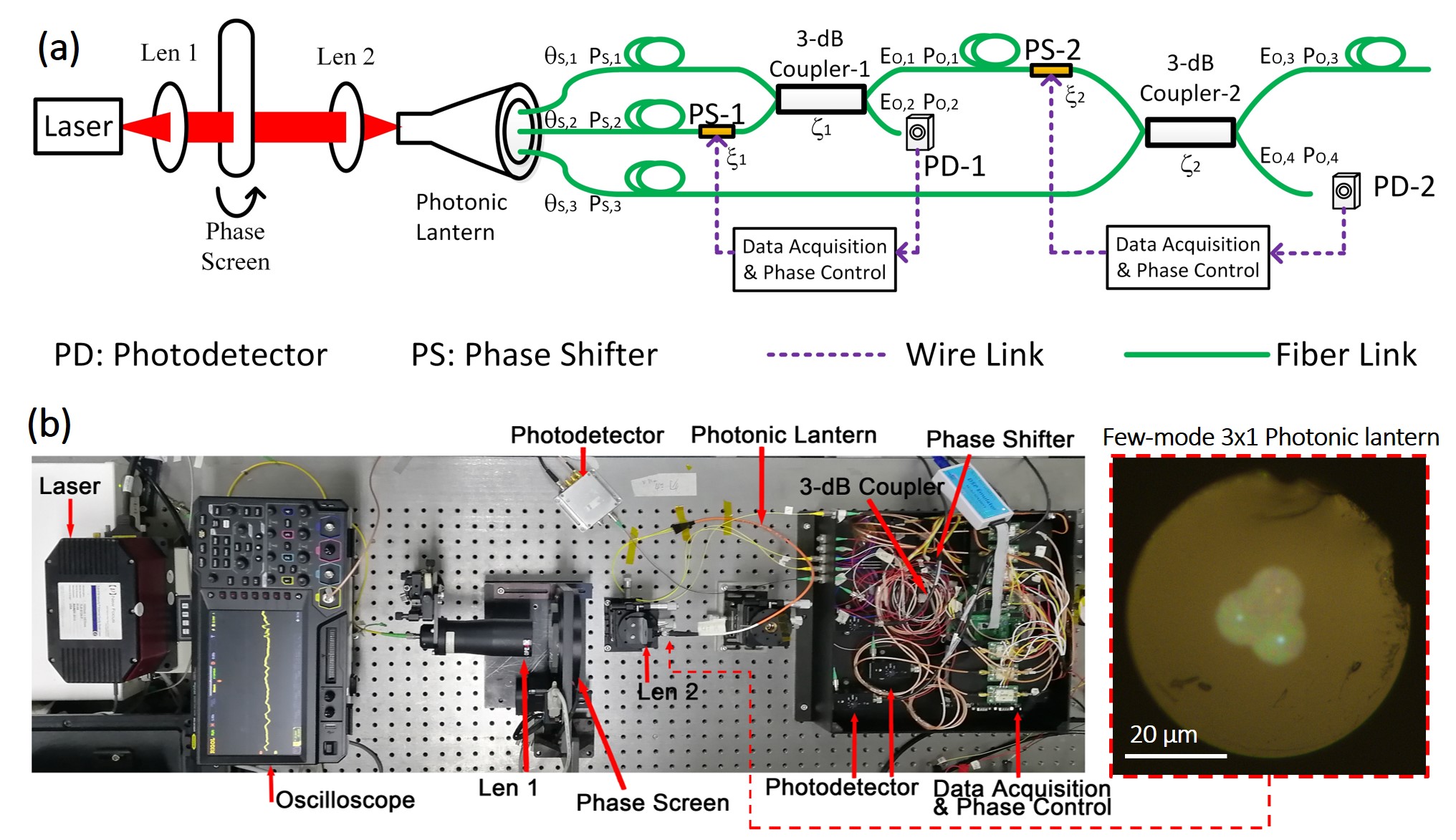}
\caption{a) Schematic diagram of the PL-based optical receiver using the method of the CBC. b) (left) Experimental configuration of the PL-based optical receiver using the method of CBC. (right) Microscope image of the few-mode end of the 3x1 PL used in the experiment.}
\label{fig: SETUP}
\end{center}
\end{figure*}

The experimental assembly of the all-fiber CBC photonic lantern optical receiver is shown in Fig. \ref{fig: SETUP}.
A 1550 nm wavelength laser was used as the source.
The emitting lens (Lens 1) had a diameter of 5 cm, and the receiving lens (Lens 2) of 5 mm. The spot diameter of the laser on the receiving lens was 3 cm, which is larger than the lens itself to simulate the beam divergence that would arise in applications in which the signal travels over long distances like in FSOC and laser satellite communications.   

The $3 \times 1$ photonic lantern was made with Corning SMF-28 fibers at the Sydney Astrophotonic Instrumentation Laboratory at The University of Sydney. The few-mode end of the photonic lantern (i.e. input aperture) had a 0.15 $NA $ and a core diameter of $18$ $\mu$m with $0.4$ dB insertion losses in the SMF to few-mode direction.
The 3-dB fiber couplers used were 2x2 optical fused couplers with a loss of $0.3$ dB. The phase shifters used for phase compensation were fibre-based FPS-002 from General Photonics and had an insertion loss of $1.5$ dB with a maximum phase compensation value of $65 \pi$ at 1550 nm.
Thorlabs APD430C photodiodes were used, and the central processing unit of the digital signal processing module used for data acquisition and phase control was a Texas Instruments model TMS320C28346. The analog to digital converter unit of the digital signal processing module was an Analog Devices model AD7606, and the digital to analog converter unit of the digital signal processing module was the AD5344 model.

\subsection{Without turbulence}

\begin{figure}
\begin{center}
\includegraphics [width=0.45\textwidth, draft=false] {{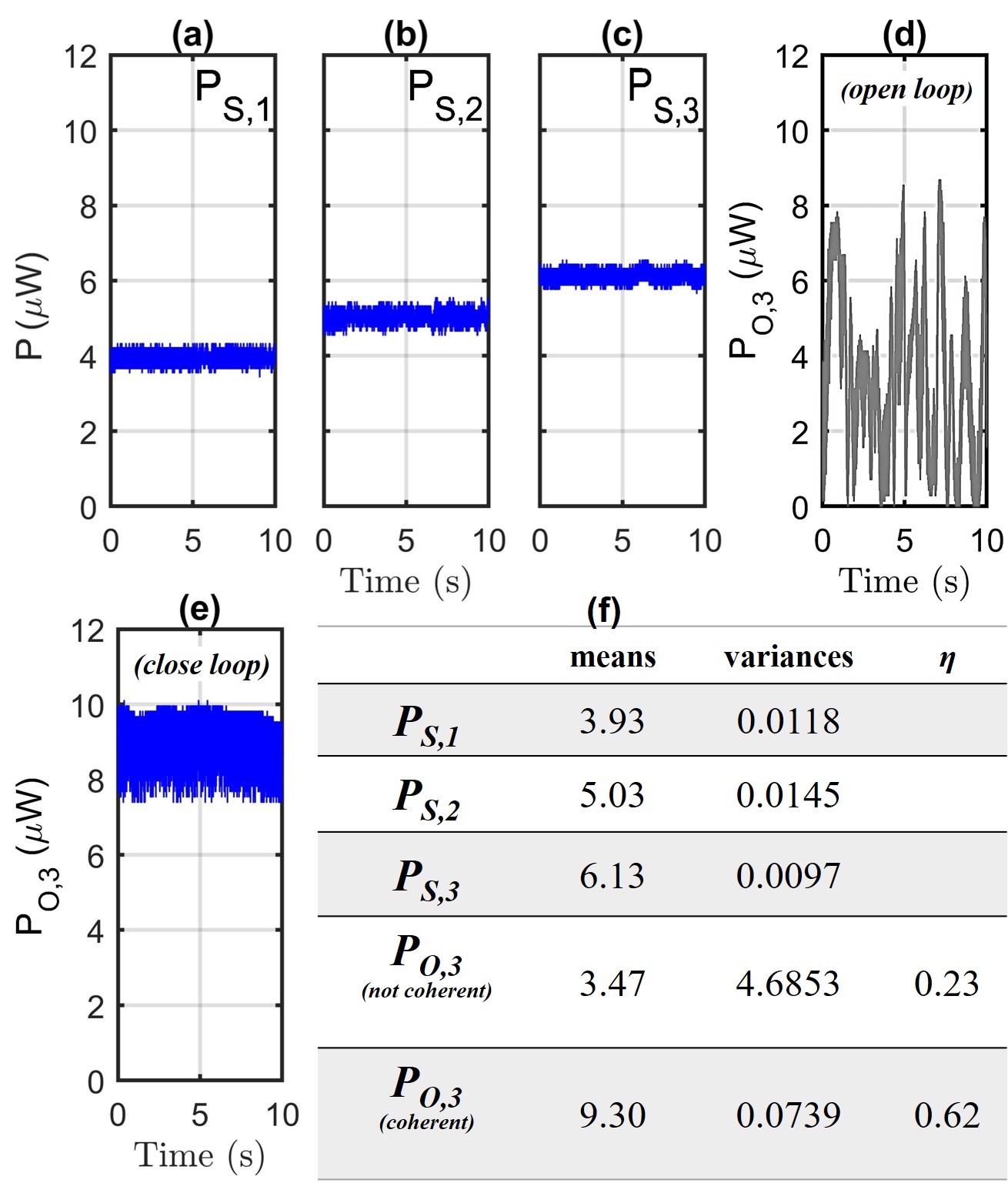}}
\caption{Experimental results of the optical powers in the absence of turbulence (a) $P_{S, 1}$; (b) $P_{S, 2}$; (c) $P_{S, 3}$; (d) ${{P}_{O, 3}}$ (open loop - beams combined but not coherently); (e) ${{P}_{O, 3}}$ (closed loop - beams coherently combined); (f) Means and variances of the optical powers in the case of atmospheric turbulence and combining efficiency.}
\label{fig: NOTUR}
\end{center}
\end{figure}

The performance of the photonic lantern optical receiver was first verified in the absence of atmospheric turbulence, and the results of the optical power outputs  are shown in Figure \ref{fig: NOTUR}. 
When the lantern is kept stationary and there is no atmospheric turbulence, the optical powers of the single-mode outputs do not change \cite{Birks2015} as expected. When the beams were combined without phase compensation (i.e. not coherently) the final average optical power was $P_{O, 3}=3.47$~ $\upmu$W with a variance of 4.6853 as shown in Figure \ref{fig: NOTUR}(d). 
This gives an efficiency of $\eta = 0.23$. When the beams were coherently combined by using the adaptive optics phase shifter controllers in a closed loop configuration, the average power increased to $P_{O, 3}=9.30$~$\upmu$W, increasing the efficiency to $\eta = 0.62$, and the variance decreased to 0.0739. This confirms the effectiveness of the all-fiber photonic lantern based optical CBC approach.

Taking the single-mode output powers of the lantern and the loss of the various components given above into account, Eqn.~\eqref{eq: effi} gives the theoretical power value of the coherently combined beams as $P_{O, 3}=10.17$~$\upmu$W. The  measured output power is 91\% of this theoretical value, we conjecture that and the discrepancy is attributed to possible mismatches in the polarisation of the combined beams (as the output of the lantern used SMF-28 fibre that is not polarisation maintaining).

In an ideal case where the insertion loss of all devices are eliminated, Eqn.~\eqref{eq: effi} gives the output values as  
$P_{O, 2}=0.03$~$\upmu$W, 
$P_{O, 3}=14.93$~$\upmu$W, 
$P_{O, 4}=0.13$~$\upmu$W, leading to a theoretical efficiency of $\eta = 0.99$.

\subsection{With turbulence}
\begin{figure}
\begin{center}
\includegraphics [width=0.45\textwidth, draft=false] {{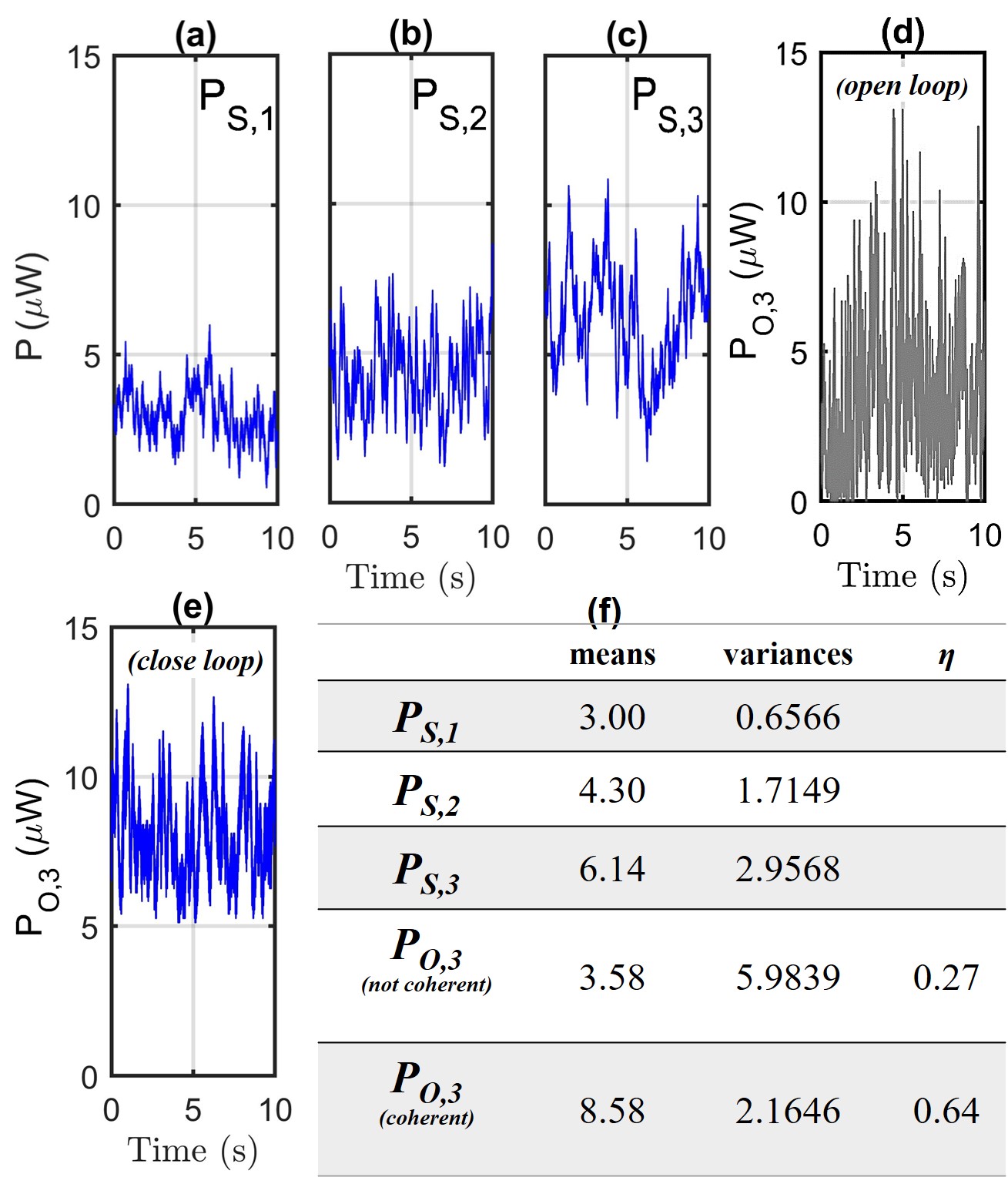}}
\caption{
Experimental results of the optical powers in the case of atmospheric turbulence.
(a) $P_{S, 1}$; (b) $P_{S, 2}$; (c) $P_{S, 3}$; (d) ${{P}_{O, 3}}$ (open loop); (e) ${{P}_{O, 3}}$ (closed loop). (f) Means and variances of the optical powers in the case of atmospheric turbulence and combining efficiency.}
\label{fig: TUR}
\end{center}
\end{figure}

We used a rotating phase screen to simulate the atmospheric turbulence in order to verify the performance of the photonic lantern receiver. The atmospheric coherent length of the phase screen is $r_{O}=3$~mm running at the Greenwood frequency of the system of $f_{G}=60$~Hz with an aperture of 5~mm (Lens 2), giving us a $D/r_{O}=1.7 $ which corresponds to weak turbulence. 
The bandwidth of the receiver and the number of modes supported on our photonic lantern proof of concept optical receiver were limited, hence we worked at the weak turbulence regime to verify the feasibility of the receiver, and the experimental results are shown in Fig. \ref{fig: TUR}. Results show that the power of the single mode outputs of the lantern fluctuated due to the variation of the input mode profile, caused by the atmospheric turbulence.
When the beams were combined without phase compensation the final average optical power was $P_{O, 3}=3.58$~$\upmu$W, with a variance of 5.9839. This gives a mean efficiency of $\eta = 0.27$. When the beams were coherently combined the average power increased to $P_{O, 3}=8.58$~$\upmu$W, increasing the mean efficiency to  $\eta = 0.64$, and the variance decreased to 2.1646. This confirms the effectiveness of our proof of concept photonic lantern CBC receiver even under conditions of atmospheric turbulence.

Taking all the known losses into account, the theoretical output is $P_{O, 3}=9.23$~$\upmu$W. The experimental value is 93\% of this amount. The ideal case where all loss is eliminated would produce $P_{O, 2}=0.06$~$\upmu$W, $P_{O, 3}=13.36$~$\upmu$W, $P_{O, 4}=0.02$~$\upmu$W, and an efficiency of $\eta = 0.99$. Although the absolute values of these results under weak turbulence are very similar to those without turbulence, the variance values are much higher showing a strong temporal response in overall power outputs as shown in Fig. \ref{fig: TUR}.

\section{Analysis of the SNR performance}
\label{SNR analysis}
An essential figure of merit to evaluate in any optical receiver for FSOC and laser satellite communications is the signal to noise ratio of the system. Here we consider the SNR of the receiver under various arrangements and a binary phase-shift keying (BPSK) modulation, which is a two-phase modulation scheme.

\subsection{Optical and Digital coherent beam combining SNR derivation}
In the CBC method used here, the final combined single-mode signal beam with power ${{P}_{O, 3}}$ is mixed with the single-mode local oscillator beam with power ${{P}_{LO}}$ in the SMF mixer \cite{Jacob1995}. The beam, after optical mixing, undergoes photoelectric conversion through the balanced photodetector. 
For simplicity in our evaluation model we assume that the dark current of the balanced photodetector is negligible and that ${{{P}_{LO}}}\gg {{P}_{O, 3}}$. Under these conditions the shot noise should become the dominant source of noise, and other sources of noise can be ignored. When a heterodyne detection architecture is used, the SNR of the this optical domain coherent beam combining (CBC) arrangement ${\rm SNR}_\text{CBC}$ can be estimated by 
\begin{equation}
{\rm SNR}_\text{CBC} =\frac{R{{P}_{O, 3}}}{qB}
\label{eq: SNRCBC}
\end{equation}

\noindent 
where $R$ is the responsivity of the photodiode, $q$ is the electronic charge and $B$ the noise equivalent bandwidth of the detector.

In the DCC case of separately measuring the single-mode outputs of the lantern and combining the signals electronically, the signal beam at the $i$th SMF end of the lantern has power ${{P}_{S,  i}}$ and  is mixed with the $i$th local oscillator beam in the SMF mixer \cite{Jacob1995}. The total oscillator beam power ${{P}_{LO}}$ is divided equally amongst the 3 outputs of the lantern and each beam after optical mixing undergoes photoelectric conversion through balanced photodetectors. 
We assume, as above, that the shot noise is the dominant source of noise.

In this scheme, the electrical signals from all 3 photodetectors are summed with equal weight. When heterodyne detection is used, the SNR of this electrical combining using the DCC method ${\rm SNR}_\text{DDC}$ is given by\cite{BOZHANG2019}
\begin{equation}
{\rm SNR}_\text{DCC} =\frac{R[{ \sqrt{{P}_{S, 1}}  +\sqrt{{P}_{S, 2}}+ \sqrt{{P}_{S, 3}} }]^2}{3qB}.
\label{eq: SNRDCC}
\end{equation}

\subsection{SNR comparison and discussion of results}

Considering an ideal case of optical combining, with an efficiency of $\eta=100 \%$ gives ${P}_{O, 3}={{{P}_{S, 1}}+{{P}_{S, 2}}+{{P}_{S, 3}}}$, and
\begin{equation}
{\rm SNR}_\text{CBC} =\frac{R[{{{P}_{S, 1}}+{{P}_{S, 2}}+{{P}_{S, 3}}}]}{qB}
\label{eq: SNRCBC2}
\end{equation}

\noindent
in the present case of three outputs. Generalising to $N$ outputs gives

\begin{equation}
{\rm SNR}_\text{CBC} =\frac{R[{{{P}_{S, 1}}+{{P}_{S, 2}}+\cdots +{{P}_{S, N}}}]}{qB}
\label{eq: SNRCBC3}
\end{equation}
for optical combining and 
\begin{equation}
{\rm SNR}_\text{DCC} =\frac{R[{ \sqrt{{P}_{S, 1}}  +\sqrt{{P}_{S, 2}}+\cdots + \sqrt{{P}_{S, N}} }]^2}{NqB}.
\label{eq: SNRDCC3}
\end{equation}
for the electrical combining. 

\begin{figure*}
\centering
\includegraphics [width=0.75\textwidth, draft=false] {{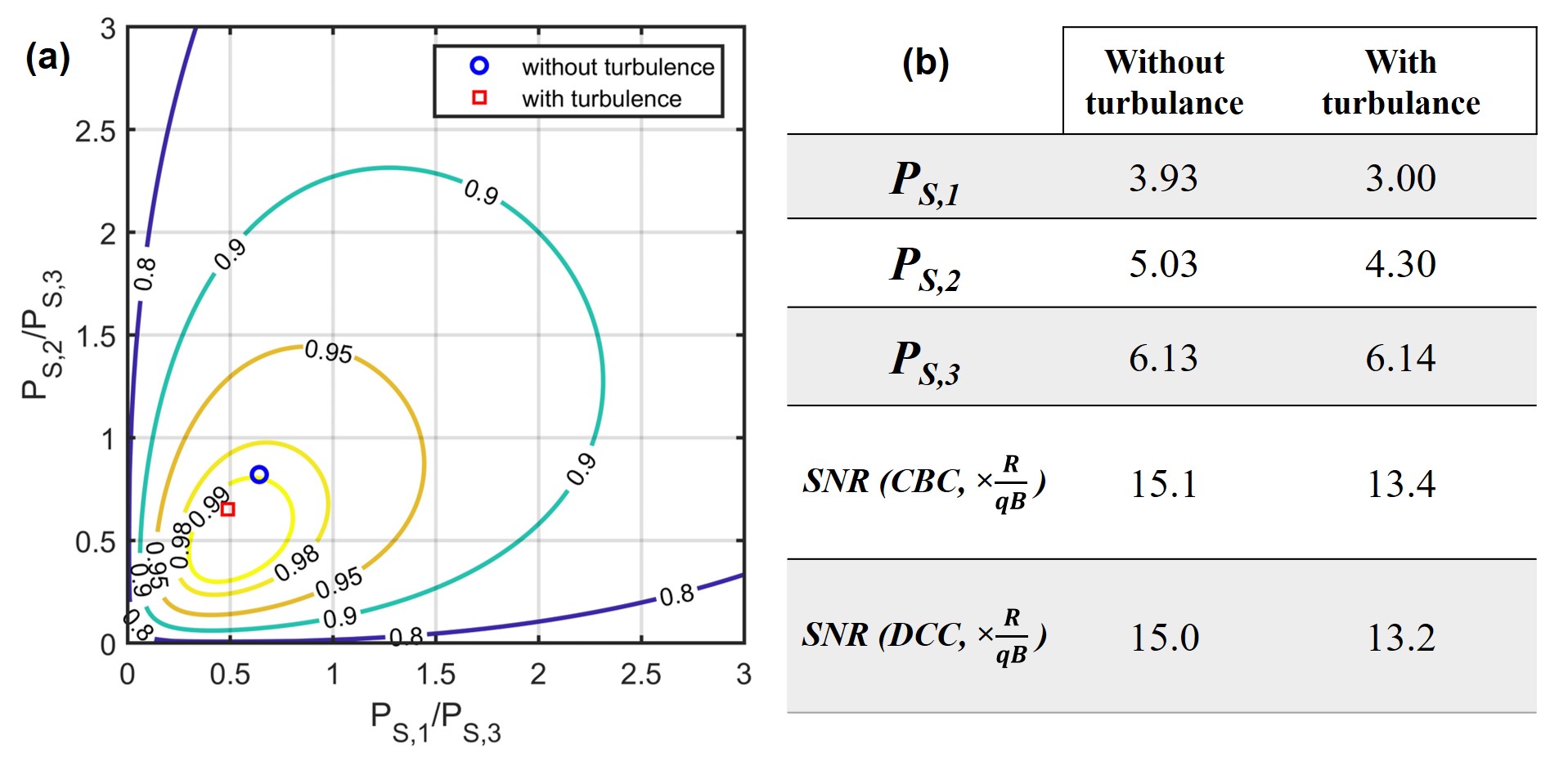}}
\caption{
(a) The relationship between the coherent synthesis efficiency and $P_{S, 1}$, $P_{S, 2}$, $P_{S, 3}$. (b) Calculated SNR from the outputs of the receiver with and without turbulence, considered using optical and electrical combining.}
\label{fig: coherence_SNR}
\end{figure*}

Equations ~\eqref{eq: SNRCBC3} and ~\eqref{eq: SNRDCC3} show that ${\rm SNR}_\text{CBC} \geq {\rm SNR}_\text{DCC}$. If the input is already diffraction limited (i.e. single-mode, in the absence of atmospheric turbulence), the electrical combining can reduce SNR whereas the optical combining approach will not offer a significant advantage in the case of a stable unperturbed diffraction-limited wavefront. Furthermore, equations \eqref{eq: SNRCBC3} and \eqref{eq: SNRDCC3} can be applied to calculate the SNR for the experimental results obtained with and without turbulence, shown in Figures \ref{fig: NOTUR} and  \ref{fig: TUR}. The comparison shown in 
Fig. \ref{fig: coherence_SNR}(b) assumes an ideal case of no power loss, and shows that the optically combined beams result in a higher SNR as expected. Figure~\ref{fig: coherence_SNR}(a) shows how the efficiency changes as a function of the relative strength of $P_{S, 1}$, $P_{S, 2}$, and $P_{S, 3}$, as calculated using Eqn.~\eqref{eq: effi} assuming all losses are eliminated. The idealised experimental results with/without turbulence are also shown.
We see that when $P_{S, 1} / P_{S, 3}<2.5$ and $P_{S, 2} / P_{S, 3}<2.5$, 
as long as $P_{S, 1}$, $P_{S, 2}$ and $P_{S, 3}$ are not very large or very small, the coherence efficiency is easily over 90\%.
When $P_{S, 1} / P_{S, 3} \approx 0.5$ and $P_{S, 2} / P_{S, 3} \approx 0.5$, 
the coherence efficiency can reach 99\%.

\begin{figure*}[t]
\begin{center}
\includegraphics [width=0.95\textwidth, draft=false] {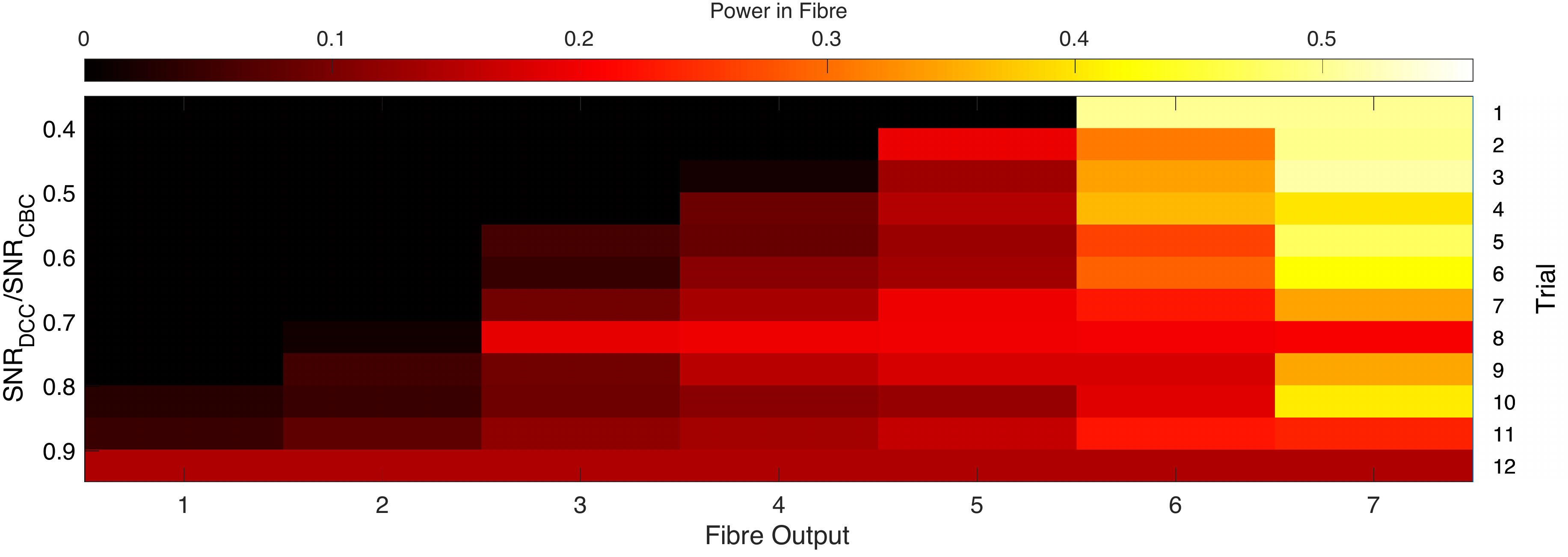}
\caption{Illustration of difference in the SNR between DCC and CBC depending on the power distribution in the output ports of the PL. Trial 1 and 12 show two special cases where all the light is in two fibres and all the light is evenly distributed between the fibres respectively. In the case of trial 12, the SNR is identical for both DCC and CBC. In all other cases the SNR for the CBC case is greater than the SNR for the DCC case. The diagram is generated using Eqn.~\eqref{eq: SNRCBC3} and Eqn.~\eqref{eq: SNRDCC3}.}
\label{fig: SNRComp}
\end{center}
\end{figure*}

In our 3 x 1 photonic lantern optical coherent receiver case study, the SNR improvement is moderate and the coherent efficiency greatly determined by the single-mode power distributions. This can be explained by the limited number of modes and single-mode outputs in our proof of concept case. It is understood, that the power distribution at the single-mode end of the lantern varies according to the input mode profile \cite{Birks2015}. Furthermore, particularly in the case of turbulence, for the electronic combination scheme the ${\rm SNR}_\text{DDC}$ is calculated based on average values of the single-mode output powers. This will mask the issue of data lost due to poor SNR variations over time. 
Hence, an additional consideration to the SNR evaluation between the optical combination and the electronic combination architectures is the effect of different output power distribution amongst the single-mode end of the lantern on the overall SNR and efficiency of the coherent combination. We see in Eqn.~\eqref{eq: SNRCBC3} and Eqn.~\eqref{eq: SNRDCC3} that the SNR for the DCC case is always lower than for the CBC apart from when the power at the lantern outputs is equal. However, this is highly unlikely to occur in a real case scenario and is not observed in the present results, as seen in 
Fig. \ref{fig: coherence_SNR}b. To illustrate and evaluate this point, Fig.~\ref{fig: SNRComp} shows an analysis for a lantern with seven outputs (7 x 1 photonic lantern). In the case where the power is distributed equally amongst the outputs (corresponding to Trial 12), the DCC and CBC approaches give the same SNR (a ratio of $SNR_{DCC}/SNR_{CBC} = 1$). However, if all the power is concentrated equally into two of the seven outputs (corresponding to Trial 1), then SNR of the DCC case reduces to only 40\% of the CBC case.  This illustrates the superiority of the CBC approach, particularly under turbulent conditions when power between the lantern's outputs is expected to fluctuate over time.

\section{Conclusion}
We proposed a photonic lantern-based multimode optical receiver based on coherent beam combining in optical fibers. Compared to existing DCC technology, CBC results in a higher SNR under realistic conditions of turbulence.
We verified the feasibility of the CBC receiver using a lantern with only three outputs as a proof of concept, active phase locking and 3-dB fiber couplers.
The experimental results showed that the coherent synthesis efficiency is $62 \%$.
Also, we used a phase screen to simulate atmospheric turbulence to verify the performance of the optical receiver and found that the coherent synthesis efficiency is $64 \%$ under atmospheric turbulence. Furthermore, the receiver performance can be further improved by reducing device losses in the system and optical components. In an ideal case of negligible device losses, the coherent synthesis efficiency could be as high as $99 \%$ in both the presence and absence of turbulence.
A theoretical analysis and numerical modelling of the SNR in a 7 x 1 photonic lantern showed that the power distribution across the single-mode outputs of the lantern played an important role. The SNR of the CBC approach remains constant whereas the SNR of the DCC approach decreases as the difference in power amongst the single-mode outputs increases. This is indeed an important factor in the presence of atmospheric turbulence and tip/tilt errors where the output of the single-mode fibers of the lantern will vary significantly with the changes in input beam. By taking this approach of active phase control in the optical combining system to increase SNR, this receiver represents a very versatile in-fibre approach to adaptive optics, with applications in free space optical communications and LIDAR.

\section*{Acknowledgement}
Bo Zhang acknowledges the support from the UCAS Joint PhD Training Program.

\normalem
\bibliographystyle{ieeetr}
\bibliography{bibfile}

\end{document}